\documentclass[preprint,aps,floatfix]{revtex4-1}
\usepackage{hyperref}
\everymath={\displaystyle}
\usepackage{young}
\usepackage{rotating}
\usepackage{longtable}
\def\1{\mbox{l\hspace{-0.53em}1}}

\usepackage{multirow}

\newlength{\AccoHaut}

\begin{document}
\title{Excited hyperons of the $N = 2$ band in the  $1/N_c$ expansion}

\author{Fl. Stancu\footnote{E-mail address: fstancu@ulg.ac.be}}
\affiliation{University of Li\`ege, Institute of Physics B5, Sart Tilman,
B-4000 Li\`ege 1, Belgium}

\date{\today}

\begin{abstract}
The  spectrum of excited baryons  in the  $N$ = 2 band is reanalyzed
in the $1/N_c$ expansion method, with  emphasis on hyperons. 
Predictions are made for the classification of these excited baryons into SU(3) singlets, octets and decuplets.
\end{abstract}

\maketitle

\section{Introduction}

The $1/N_c$ expansion method \cite{'tHooft:1973jz,Witten:1979kh} 
where $N_c$ is the number of colors, is a powerful and  systematic tool
for baryon spectroscopy.
For $N_f$ flavors, the ground state baryons
display an exact contracted SU($2N_f$) spin-flavor symmetry in the
large $N_c$ limit of QCD \cite{Gervais:1983wq,DM93}. 
The Skyrme model, the strong coupling theory  and the static quark model share a common
underlying symmetry with QCD baryons in the large $N_c$ limit \cite{Bardakci:1983ev}.

The method  has been successfully applied 
to ground state baryons ($N = 0$ band), in the
symmetric representation $\bf 56$ of SU(6) 
\cite{DM93,Jenk1,DJM94,DJM95,JL95}. 
At $N_c \rightarrow \infty$ the ground state baryons are degenerate.
At large, but finite $N_c$, the mass splitting 
starts at order $1/N_c$ as first observed in Ref. \cite{Bardakci:1983ev}.

The extension of the $1/N_c$ expansion method to excited states 
requires the symmetry group  SU($2N_f$) $\times$ O(3) \cite{Goi97}, in order to introduce orbital excitations.
It happens that the experimentally observed resonances can approximately be classified  
as  SU($2N_f$) $\times$ O(3) multiplets, grouped into 
excitation bands, $N $ = 1, 2, 3, ..., 
each band containing a number of SU(6) $\times$ O(3) multiplets.

The situation is technically more complicated for mixed symmetric states than for symmetric states. Two approaches 
have been proposed so far. The first one is based on the Hartree approximation and
describes the $N_c$ quark system as a ground state symmetric core of $N_c - 1 $ 
quarks and an excited quark \cite{CCGL}. 

The second procedure, where the Pauli principle is
implemented to all $N_c$ identical quarks has been proposed in Refs. \cite{Matagne:2006dj,Matagne:2008kb}.
There is no physical reason to separate the excited quark from the rest of the system.
The method can straightforwardly be applied to all excitation bands. It requires 
the knowledge of the matrix elements of all the  SU($2N_f$) generators acting on mixed 
symmetric states described by the partition $(N_c - 1,1) $.
In both cases the mass splitting starts at order $N^0_c$. 
The latest achievements for the ground state and the current
status of large $N_c$ QCD excited 
baryons ($N$ = 1, 2, 3, 4) can be found in Ref. \cite{Matagne:2014lla}.
The $N$ = 1 band is the most studied. The $N$ = 2 band received considerable attention too.
Here we reanalyze the results of Ref.  \cite{Matagne:2016gdc}  for $N$ = 2. The reason is that
in a few octets an anomalous situation appeared where the hyperons $\Lambda$ or $\Sigma$
(presently degenerate) appeared slightly lighter than the nucleon in the same octet. 

Here we use the data of the 2014 Particle Data Group \cite{PDG} which includes
changes due to a more complex analysis of all major photo-production of mesons in a coupled-channel   
partial wave analysis.

%%%%%%%%%%%%%%%%%%%%%%%%%%%%%%%%%%%%%%%%%%%%%%%%%%%%%%%%%%%%%%%%%%%%%%%%%%%
\section{The Mass Operator}\label{massformula}

The general form of the mass operator,  where the SU(3) symmetry is broken, has first been proposed in Ref. \cite{JL95} as
\begin{equation}
\label{massoperator}
M = \sum_{i}c_i O_i + \sum_{i}d_i B_i .
\end{equation} 
The operators $O_i$ are defined as the scalar products
\begin{equation}\label{OLFS}
O_i = \frac{1}{N^{n-1}_c} O^{(k)}_{\ell} \cdot O^{(k)}_{SF},
\end{equation}
where  $O^{(k)}_{\ell}$ is a $k$-rank tensor in SO(3) and  $O^{(k)}_{SF}$
a $k$-rank tensor in SU(2)-spin, but invariant in SU($N_f$).
Thus $O_i$ is rotational invariant.
For the ground state one has $k = 0$. The excited
states also require  $k = 1$  and $k = 2$ terms.
The $k = 1$ tensor has three components, which are the generators $L^i$ of SO(3). 
The components of the $k = 2$ tensor operator of SO(3) read %cite{Matagne:2005gd}
\begin{equation}\label{TENSOR} 
L^{(2)ij} = \frac{1}{2}\left\{L^i,L^j\right\}-\frac{1}{3}(-)^i
\delta_{i,-j}\vec{L}\cdot\vec{L}.
\end{equation}
The operators $O^{(k)}_{SF}$ are expressed in terms of the SU($N_f$) generators
$S^i$, $T^a$ and $G^{ia}$.

The operators $B_i$ break the SU(3) flavor symmetry and are defined to have zero expectation
values for nonstrange baryons. 
The coefficients $c_i$ encode the quark dynamics and $d_i$ measure the SU(3) breaking.
They are obtained from a numerical fit.
The most dominant operators considered in the mass formula together with the fitted
coefficients  are presented in Table \ref{operators}.

For the $[{\bf 56}]$-plets the spin-orbit operator $O_2$  
is defined in terms of angular momentum $L^i$ components acting on the whole
system as in Ref. \cite{GSS03} and is order  $\mathcal{O}(1/N^c)$ 
\begin{equation}\label{LS}
 O_2 = \frac{1}{N_c} L \cdot S, ~~~   
\end{equation}
while for the $[{\bf 70}]$-plets it is defined as a single-particle operator $\ell \cdot s$ of order $\mathcal{O}(N^0_c)$. 
\begin{equation}\label{newspinorbit}
O_2 = \ell \cdot s = \sum^{N_c}_{i=1} \ell(i) \cdot s(i).
\end{equation}

%%%%%%%%%%%%%%%%%%%%%%%%%%%%%%%%%%%%%%%%%%%%%%%%%%%%%%%%%%%%%%%%%%%%%%%%%%%%%

\section{Matrix elements}\label{matrixelements}

The matrix elements of the $[\bf 56,2^+]$ multiplet were derived in Ref. \cite{GSS03}.
Details of the derivation of the matrix elements of $O_i$ for $[\bf 70,\ell^+]$,
as a function of $N_c$, can be found in Ref. \cite{Matagne:2013cca}.
Note that in the case of mixed symmetric states  the matrix elements of  $O_6$ are  $\mathcal{O}(N^0_c)$,
in contrast to the symmetric case where they are $\mathcal{O}(N^{-1}_c)$, and  non-vanishing only
for octets, while for the symmetric case they are non-vanishing for decuplets. 
Thus, at large $N_c$ the splitting starts at order $\mathcal{O}(N^{0}_c)$ for mixed symmetric states 
due both to $O_2$ and $O_6$.

The SU(3) flavor breaking operators $B_i$ have the same definition for both the symmetric and mixed symmetric multiplets. 
The matrix  elements of $B_2$ and $B_3$  for $[\bf 70,\ell^+]$ were first calculated in Ref.  \cite{Matagne:2016gdc}. 
For practical purposes we have summarized
these results by two simple analytic formulas valid at $N_c$ = 3. 
The diagonal matrix elements of $B_2$ take the following form
\begin{equation}\label{B2}
  B_2 = - n_s 
%\mathcal{S} 
\frac{\langle L \cdot S \rangle}{6 \sqrt{3}},
\end{equation}
where $n_s$ is the number of strange quarks and
${\langle L \cdot S \rangle}$ is the expectation value of the spin-orbit operator acting on the whole system.
Similarly 
the diagonal matrix elements of $B_3$ take the simple analytic form 
\begin{equation}\label{B3}
  B_3  = - n_s
\frac{S(S + 1)}{6 \sqrt{3}},
\end{equation}
where  $S$ is the total spin. The contribution of $B_3$ is always negative, otherwise vanishing 
for nonstrange baryons.
These formulas can be applied to  $^28_J$, $^48_J$, 
$^2{10}_J$ and $^2{1}_{1/2}$ baryons of the  $[{\bf 70},\ell^+]$ multiplet. 
Presently the SU(3) breaking operators $B_2$ and $B_3$ are included  in the analysis
of the   $[\bf 70,\ell^+]$ multiplet, first considered in  Ref. \cite{Matagne:2016gdc}.

%%%%%%%%%%%%%%%%%%%%%%%%%%%%%%%%%%%%%%%%%%%%%%%%%%%%%%%%%%%%%%%%%%%%%%%%%
\section{Fit and discussion}\label{fit}

We have performed a consistent  analysis of the experimentally known 
resonances supposed to belong either to the symmetric  %$[56',0+]$
$[{\bf 56},2^+]$ multiplet or 
to the mixed symmetric multiplet $[{\bf 70},\ell^+]$ with $\ell$ = 0 or 2,
by using the same operator basis. Results of the fitted coefficients $c_i$ and
$d_i$ are exhibited in  Table \ref{operators} together with the values of $\chi_{\mathrm{dof}}^2$
for each multiplet.

The spin and flavor operators $O_3$ and $O_4$ are the dominant two-body operators and bring important
$1/N_c$ corrections to the masses.
The sum of $c_3$ and $c_4$ of  $[{\bf 70},\ell^+]$ is comparable to the value  of $c_3$
in $[56,2^+]$ where the equal contribution of $O_3$ and $O_4$ is included in $c_3$. 
The contribution of the operator $O_6$ containing an SO(3) tensor is important especially for  $[{\bf 70},\ell^+]$ multiplet.
Together with the spin-orbit it may lead  to the mixing of doublets and quartets to be considered in further  studies when the accuracy of 
data will increase. The incorporation of $B_2$ and $B_3$ in the mass formula of the $[{\bf 70},\ell^+]$ multiplet
brings more insight into  the SU(6) multiplet classification of excited baryons in the $N$ = 2 band.

%%%%%%%%%%%%%%%%%%%%%%%%%%%%%%%%%%%%%%%%%%%%%%%%%%%%%%%%%%%%%%%%%%%%
\begin{table}[htb]
\caption{List of the dominant operators and their coefficients (MeV) from the mass formula (\ref{massoperator}) obtained 
in numerical fit for  $[{\bf 56},2^+]$ in column 2 and for $[{\bf 70},\ell^+]$ in column 3.
The spin-orbit operator $O_2$ is defined by Eq. (\ref{LS}) for $[{\bf 56},2^+]$ and by Eq.(\ref{newspinorbit})
for $[{\bf 70},\ell^+]$. }
\label{operators}
\renewcommand{\arraystretch}{1.2} % enlarge line spacing
\begin{tabular}{lrrrrr}
\hline
\hline
Operator & $[{\bf 56},2^+]$ &\hspace{0.5cm} &  $[{\bf 70},\ell^+]$ \hspace{0.01cm}  &\\
\hline
\hline
$O_1 = N_c \ \1 $                               &   542 $\pm$ 2    & &   631 $\pm$ 10 &    \\
$O_2$ spin-orbit                  &                 7 $\pm$10      & &   62 $\pm$ 26 &    \\
$O_3 = \frac{1}{N_c}S^iS^i$                     &   233 $\pm$ 11   & &    91 $\pm$ 31 &     \\
$O_4 = \frac{1}{N_c}\left[T^aT^a-\frac{1}{12}N_c(N_c+6)\right]$ &  & &   112 $\pm$ 22 &   \\[0.8ex]
%$O_5 =  \frac{3}{N_c} L^{i} T^{a} G^{i}$           & -22 $\pm$ 5  & &             & &\\ 
$O_6 =  \frac{1}{N_c} L^{(2)ij} G^{ia} G^{ja}$ &   6 $\pm 19$      & &    137 $\pm$ 55 &    \\[0.5ex]
\hline
$B_1 = n_s$                                     &  205 $\pm$ 14    & &     35 $\pm$ 33 &      \\ 
$B_2 = \frac{1}{N_c}(L^iG^{i8}  - \frac{1}{2\sqrt{3}} L^i S^i)$    &    97 $\pm$ 40     & & - 38 $\pm$ 121 & \\
$B_3 = \frac{1}{N_c}(S^iG^{i8}  - \frac{1}{2\sqrt{3}} S^i S^i)$    & 197 $\pm$ 69       & &   46 $\pm$ 159 &  \\[0.8ex]
\hline                  
$\chi_{\mathrm{dof}}^2$                                            &     1.63           & &    1.67 &    \\
\hline \hline
\end{tabular}
\end{table}

%%%%%%%%%%%%%%%%%%%%%%%%%%%%%%%%%%%%%%%%%%%%%%%%%%%%
\subsection{The multiplet $[{\bf 56},2^+]$}

The partial contribution and the calculated total mass obtained from the fit were presented in Table VI of
Ref. \cite{Matagne:2016gdc} which we do not repeat here.
The experimental masses were taken from the 2014 version of the Review of Particle Properties (PDG) \cite{PDG},
except for $\Delta(1905)5/2^+$ where we used the mass of Ref. \cite{GSS03} which gives a smaller $\chi_{\mathrm{dof}}^2$,
but does not much change the fitted values of $c_i$ and $d_i$. As expected, the most important sub-leading contribution comes 
from the spin operator $O_3$. The contributions of the angular momentum-dependent operators $O_2$ and $O_6$ are 
comparable, but small. Among the SU(3) breaking terms, $B_1$ is dominant. An important remark is that 
in the  $[{\bf 56},2^+]$ multiplet $B_2$ and $B_3$ lift the degeneracy of $\Lambda$ and $\Sigma$ baryons in 
the octets, which is not the case for the $[{\bf 70},\ell^+]$ multiplet.

%%%%%%%%%%%%%%%%%%%%%%%%%%%%%%%%%%%%%%%%%%%%%%%%%%%%%%%%%%%%%%%%%ù

\subsection{The multiplet $[{\bf 70},\ell^+]$}

As compared to Ref. \cite{Matagne:2013cca} where only 11 resonances have been included in the numerical fit, here
we consider 16 resonances, having a status of three, two or one star.
This means that we have tentatively added  the resonances
$\Xi(2120)?^{?*}$, $\Sigma(2070)5/2^{+*}$, $\Sigma(1940)?^{?*}$,  $\Xi(1950)?^{?***}$ and
$\Sigma(2080)3/2^{+**}$. The masses and the error bars considered
in the fit correspond to averages over data from the particle listings,
except for a few which favor specific experimental values cited in the headings of Table \ref{MASSES}.

We have ignored the $N(1710){1/2^{+***}}$ and the 
$\Sigma(1770){1/2^{+*}}$ resonances, 
the theoretical argument being that their masses are too low, leading to unnatural sizes for the 
coefficients  $c_i$ or $d_i$ \cite{Pirjol:2003ye}.
On the experimental side one can justify the removal of  the controversial $N(1710)1/2^{+***}$ resonance due to
the latest GWU analysis  of Arndt et al. \cite{Arndt:2006bf} where it has not been seen. 
We have also ignored
the $\Delta(1750)1/2^{+*}$ resonance, because
neither Arndt et al. \cite{Arndt:2006bf} nor Anisovich et al.  \cite{Anisovich:2011fc} find evidence
for it.

%%%%%%%%%%%%%%%%%%%%%%%%%%%%%%%%%%%%%%%%%%%%%%%%%%%%%%
{\squeezetable
\begin{table}
%\begin{sidewaystable}
\caption{Partial contribution and the total mass (MeV) predicted by the $1/N_c$ expansion. 
%using Fit 2 
%with operators of Tables \ref{BARYON70} and \ref{break70}.
The last two columns give  the empirically known masses and status from the 2014 Review of Particles Properties 
\cite{PDG} unless specified by (A) from \cite{Anisovich:2011fc}, 
(L) from \cite{Litchfield:1971ri},
(Z) from \cite{Zhang:2013sva},
(G1) from  \cite{Gopal:1980ur}, (B) from \cite{Biagi:1986vs},
(AB) from \cite{Ablikim:2009iw},
(G2) from  \cite{Gopal:1976gs},
%  (B) from Ref. \cite{Berthon:1970tg},(S) from Ref. \cite{Shrestha:2012ep}
.}\label{MASSES}
\renewcommand{\arraystretch}{1.5}
\begin{tabular}{crrrrrrcccccl}\hline \hline
                    &      \multicolumn{8}{c}{Part. contrib. (MeV)}  & \hspace{0.0cm} Total (MeV)   & \hspace{.0cm}  Experiment (MeV)\hspace{0.0cm}
 &\hspace{0.cm}  Name, status \hspace{.0cm} \\

\cline{2-9}
                    & \hspace{.0cm} $c_1O_1$  & \hspace{.0cm}  $c_2O_2$ & \hspace{.0cm}$c_3O_3$ &\hspace{.0cm}  $c_4O_4$ 
&\hspace{.0cm}  $c_6O_6$ & $d_1B_1$ & $d_2B_2$ & $d_3B_3$ &  \\
\hline
$^4N[{\bf 70},2^+]\frac{7}{2}^+$        & 1892 & 62 & 113 & 28 & - 22 & 0  &  0  &    0  & $2073\pm 38$  & $2060\pm65$(A) & $N(1990)7/2^+$**  \\
$^4\Lambda[{\bf 70},2^+]\frac{7}{2}^+$  &       &    &     &    &     & 35 & 11  & - 17  & $2102\pm 19$  & $2100\pm30$(L) 
& $\Lambda(2020)7/2^+$* \\
%$^4\Sigma[{\bf 70},2^+]\frac{7}{2}$   &   & 1 & 5/4 & 1/4 & -5/2 & 1  & - $\frac{\sqrt{3}}{6}$ & - $\frac{5\sqrt{3}}{24}$  &   
%&     & $\Sigma(2030)7/2^+$**** \vspace{0.2cm}\\
$^4\Xi[{\bf 70},2^+]\frac{7}{2}^+$      &       &    &     &    &      & 70 & 22  & - 34 & $2131\pm 8$ & 
$2130\pm8$  &  $\Xi(2120)?^?$* \vspace{0.2cm}\\
\hline
%%%%%%%%%%%%%%%%%%%%%%%%%%%%%%%%%%%%%%%
$^4N[{\bf 70},2^+]\frac{5}{2}^+$        & 1892 & - 10  & 113 & 28 & 57 & 0  & 0   & 0    & $2080\pm32$ & $2000\pm50$ & $N(2000)5/2^+$**\\
%\vspace{0.2cm} \\
$^4\Lambda[{\bf 70},2^+]\frac{5}{2}^+$  &      &       &     &    &    & 35 & - 2 & - 17 & $2096\pm10$  & $2100\pm10$ & $\Lambda(2110)5/2^+$***\\
%$^4\Sigma[{\bf 70},2^+]\frac{5}{2}$   &   & -1/6  & 5/4  & 1/4 & 5/12 & 1  & $\frac{\sqrt{3}}{36}$ & - $\frac{5\sqrt{3}}{24}$ &  
% &  &  \vspace{0.2cm}\\
%$^4\Xi[{\bf 70},2^+]\frac{5}{2}$      &      &     &    &     &   & 48  & $2183\pm92$   &              & & \vspace{0.2cm}\\
%%%%%%%%%%%%%%%%%%%%%%%%%%%%%%%%%%%%%%%%
\hline
$^4N[{\bf 70},2^+]\frac{3}{2}^+$        & 1892 & - 62  &  113 & 28  & 0  &  0  &    0  & 0     & $1972\pm29$       &    & \vspace{0.2cm} \\
$^4\Lambda[{\bf 70},2^+]\frac{3}{2}^+$  &      &       &      &     & 0  & 35  & - 11  & - 17  & $1979\pm39$       & \\
%$^4\Sigma[{\bf 70},2^+]\frac{3}{2}$   &   & -1 & 5/4 & 1/4   & 0  &  1  & $\frac{\sqrt{3}}{6}$  & - $\frac{5\sqrt{3}}{24}$ & & &\vspace{0.2cm} \\
%$^4\Xi[{\bf 70},2^+]\frac{3}{2}$      &      &     &    &     &   & 48  & $2065\pm90$   &              & & \\
\hline
%%%%%%%%%%%%%%%%%%%%%%%%%%%%%%%%%%%%%%
$^4N[{\bf 70},2^+]\frac{1}{2}^+$        & 1892 &- 93 & 113 & 28 &- 80 & 0  & 0 & 0   & $1861\pm33 $ & $1870\pm35$(A) & $N(1880)1/2^+$**
\vspace{0.2cm}\\
$^4\Lambda[{\bf 70},2^+]\frac{1}{2}^+$  &      &      &     &    &    & 35 & -16 &- 16& $1869\pm79 $ &                & \\              
 \hline
%%%%%%%%%%%%%%%%%%%%%%%%%%%%%%%%%%%%%%
$^2N[{\bf 70},2^+]\frac{5}{2}^+$        & 1892 & 21  & 23  & 28  & 0   & 0   &  0 &  0   & $1964\pm29$ & $1860\pm{^{120}_{60}}$(A) & $N(1860)5/2^+$** \\
%$^2\Lambda[{\bf 70},2^+]\frac{5}{2}^+$  &      &     &     &     & 0   & 76  & 17 & - 20 & $2031\pm11$ &  $2036\pm13$(Z)  
%&  $\Lambda(2110)5/2^+$***           \\
$^2\Sigma[{\bf 70},2^+]\frac{5}{2}^+$   &      &     &     &      & 0   & 35  &  4 & - 3 & $2000\pm18$ &  $2051\pm25$(G1) 
& $\Sigma(2070)5/2^+$* \vspace{0.2cm}\\
%$^2\Xi[{\bf 70},2^+]\frac{5}{2}$      &      &     &    &     &     & 48  &   $1959\pm88$  &              & & \vspace{0.2cm}\\
%%%%%%%%%%%%%%%%%%%%%%%%%%%%%%%%%%%%%%%%
\hline
$^2N[{\bf 70},2^+]\frac{3}{2}^+$        & 1892 &- 31 & 23  & 28  & 0   &  0  & 0  & 0   & $1912\pm21$ & $1905\pm30$(A) & $N(1900)3/2^+$***  \\
%$^2\Lambda[{\bf 70},2^+]\frac{3}{2}$  &      &     &    &     &   & 24  & $1907\pm52$   &              & & \\
$^2\Sigma[{\bf 70},2^+]\frac{3}{2}^+$   &      &     &     &     & 0   &  35 & - 6 & - 3 & $1938\pm10$ & $1941\pm18$ & $\Sigma(1940)?^?$*  
\vspace{0.2cm} \\
$^2\Xi[{\bf 70},2^+]\frac{3}{2}^+$      &      &     &     &     & 0   &  70 & - 11 & - 7 & $1964\pm7$ & $1967\pm7$(B) & $\Xi(1950)?^?$*** \vspace{0.2cm}\\
%%%%%%%%%%%%%%%%%%%%%%%%%%%%%%%%%%%%%%%%%%%%%%%
\hline
$^4N[{\bf 70},0^+]\frac{3}{2}^+$         & 1892 &  0  & 113 & 28  &  0  &  0  & 0 & 0      & $2033\pm18$ & $2040\pm28$(AB)  & $N(2040)3/2^+$*\\
%$^4\Lambda[{\bf 70},0^+]\frac{3}{2}^+$  &      &     &     &     & 24  &  $2076\pm49$   &              & & \\
$^4\Sigma[{\bf 70},0^+]\frac{3}{2}^+$    &      &     &     &     &     &  35 & 0 & - 16  & $2052\pm21$&  $2100\pm69$   
& $\Sigma(2080)3/2^+$**\vspace{0.2cm} \\
%$^4\Xi[{\bf 70},0^+]\frac{3}{2}$      &      &     &    &     &   & 48  &  $2100\pm86$   &              & & \vspace{0.2cm}\\
%%%%%%%%%%%%%%%%%%%%%%%%%%%%%%%%%%%%%%%%%
\hline
$^2\Delta[{\bf 70},2^+]\frac{5}{2}^+$       & 1892  & - 21  & 23 & 140  & 0  &  0 & 0 &  0  & $2034\pm31$  & $1962\pm139$ & 
$\Delta(2000)5/2^+$**\vspace{0.2cm}\\
\hline
$^2\Sigma^{\ast}[{\bf 70},0^+]\frac{1}{2}^+$& 1892  &   0   & 23 & 140  & 0  & 35 & 0 & - 3 & $2087\pm30$   & $1902\pm96$  & $\Sigma(1880)1/2^+$** 
\vspace{0.2cm}\\
\hline
$^2\Lambda'[{\bf 70},0^+]\frac{1}{2}^+$     & 1890  &   0   & 23 & - 84 & 0  & 35 & 0 & - 3 & $1863\pm19$   & $1853\pm20$(G2) & $\Lambda(1810)1/2^+$*** 
\vspace{0.2cm}\\
\hline
\hline
\end{tabular}
\end{table}}
%%%%%%%%%%%%%%%%%%%%%%%%%%%%%%%%%%%

The partial contributions and the calculated total masses obtained from the fit are presented in Table  \ref{MASSES}.
Regarding the contribution of various operators we note that the good fit for $N(1880)1/2^{+**}$
was due to contribution of the spin-orbit operator $O_2$ of -$93$ MeV and of the operator 
$O_6$ which contributed with $-80$ MeV. The good fit also suggests that  $\Sigma(1940)?^{?*}$
and $\Xi(1950)?^{?***}$ assigned by us to the $^2[{\bf 70},2^+]3/2^+$ multiplet is reasonable, thus these 
resonances may have $J^P$ = $3/2^+$, to be experimentally confirmed in the future. 

The $1/N_c$ expansion is based on the SU(6) symmetry which naturally allows a classification of
excited baryons into octets, decuplets and singlets. In Table \ref{MASSES} 
the experimentally known resonances are presented. In addition some predictions 
are made for unknown resonances.    Many of the partners in a given
SU(3) multiplet are not known. Note that $\Lambda$ and $\Sigma$ are degenerate in our approach.
Although the operators $B_2$ and $B_3$ have different analytic forms 
at arbitrary $N_c$   \cite{Matagne:2016gdc}
they become identical at $N_c$ = 3
for $\Lambda$ and $\Sigma$ in octets, thus they cannot  lift the degeneracy between these 
hyperons, contrary to the $[{\bf 56},2^+]$ multiplet.

The present findings can be compared to the suggestions for assignments
in the $[{\bf 70},\ell^+]$ multiplet made in Ref. \cite{Crede:2013kia} as educated guesses. 
The assignment  of  $\Sigma(1880)1/2^{+**}$ 
as a $[{\bf 70},0^+]1/2^+$ decuplet resonance is  confirmed as well as the assignment  of $\Lambda(1810)1/2^{+***}$ 
as a flavor singlet.
We agree with Ref.  \cite{Crede:2013kia} regarding 
$\Lambda(2110)5/2^{+***}$ as a partner of $N(2000)5/2^{+**}$ in a spin quartet,
contrary to our previous work \cite{Matagne:2016gdc}
where 
$\Lambda(2110)5/2^{+***}$ was a member of a spin doublet, 
together with  $N(1860)5/2^{+**}$ and $\Sigma(2070)5/2^{+*}$.  
This helps to restore the correct hierarchy of masses in all octets.
However we  disagree with Ref. \cite{Crede:2013kia}
that $N(1900)3/2^{+***}$ is a member of a spin quartet.
We propose it as a partner of $\Sigma(1940)?^{?*}$ and $\Xi(1950)?^{?***}$
in a spin doublet.

The problem of assignment is not trivial.
Within  the  $1/N_c$ expansion method Ref. \cite{GSS03} suggests that  
$\Sigma(2080)3/2^{+**}$ and $\Sigma(2070)5/2^{+*}$ could be members 
of two distinct decuplets in the $[{\bf 56},2^+]$ multiplet. 

Here the important result is that the hierarchy of masses 
as a function of the strangeness is correct for all multiplets. 
An extended analysis of large $N_c$ excited hyperons can
be found in Ref. \cite{Stancu:2016ycq}.

%%%%%%%%%%%%%%%%%%%%%%%%%%%%%%%%%%%%%%%%%%%%%%%%%%%%%%%%%%%%%%%%%%%%%%%%

\vspace{1cm}

\centerline{\bf Acknowledgments}

Support from the Fonds de la Recherche Scientifique - FNRS under the
Grant No. 4.4501.05 is gratefully acknowledged.

%%%%%%%%%%%%%%%%%%%%%%%%%%%%%%%%%%%%%%%%%%%%%%%%%%%%%%%%%%%

\end{document}